# Single Silicon Vacancy Centers in 10-Nanometer Diamonds for Quantum Information Applications

*Stepan V. Bolshedvorskii[§,†,#], Anton I. Zeleneev[†,‡], Vadim V.Vorobyov[§,¶], Vladimir V.Soshenko[§,#], Olga R. Rubinas[§,†,#], Leonid A. Zhulikov[†,‡], Pavel A. Pivovarov[⁂], Vadim N. Sorokin[§,‡], Andrey N.Smolyaninov[#], Liudmila F. Kulikova[⊥], Anastasiia S. Garanina[∥], Sergey G. Lyapin[⊥], Vyacheslav N. Agafonov[∥], Rustem E. Uzbekov[†], Valery A. Davydov[⊥] and Alexey V. Akimov[‰,§,‡,*]*

[§]P.N. Lebedev Physical Institute, 53 Leninskij Prospekt, Moscow, 119991 Russia

[†]Moscow Institute of Physics and Technology, 9 Institutskiy per., Dolgoprudny, Moscow Region, 141700, Russia

[#]Photonic Nano-Meta Technologies, Moscow , The territory of Skolkovo Innovation Center, Str. Nobel b.7, 143026 Russia

[‡]Russian Quantum Center, 100 Novaya St., Skolkovo, Moscow, 143025, Russia

[¶]Institute of Physics, University of Stuttgart and Institute for Quantum Science and Technology IQST, 70569, Pfaffenwaldring 57, Stuttgart, Germany

[⁂]Prokhorov General Physics Institute of Russian Academy of Sciences, 38 Vavilov str., Moscow, 119991, Russia

⊥L.F. Vereshchagin Institute for High Pressure Physics, Russian Academy of Sciences, Troitsk, Moscow, 108840, Russia

∥GREMAN, UMR CNRS CEA 6157, F. Rabelais University, 37200 Tours, France

†Faculty of Medicine, F. Rabelais University, 37032 Tours, France

‰Texas A & M University, College Station, TX 77843, USA

**Abstract**

Ultra-small (about 10 nm), low-strain, artificially produced diamonds with an internal, active color center have substantial potential for quantum information processing and biomedical applications. Thus, it is of great importance to be able to artificially produce such diamonds. Here, we report on the high-pressure, high-temperature synthesis of such nanodiamonds about 10 nm in size and containing an optically active, single silicon-vacancy color center. Using special sample preparation technique, we were able to prepare samples containing single nanodiamonds on the surface. By correlating atomic-force microscope images and confocal optical images we verified presents of optically active color centers in single nanocrystals, and using second-order correlation measurements proved single-photon emission statistics of this nanodiamonds. This color centers have non-blinking, spectrally narrow emission with narrow distribution of spectral width and positions of zero-phonon line thus proving high quality of the nanodiamonds produced.

**Introduction**

Single-photon sources are of great interest for various applications in quantum information,[1] and in particular, quantum cryptography.[2] In addition, color centers in diamond attract a lot of attention due to the possibility of generating single photons at room temperature, low spread in spectral characteristics and access to electron spin.[3,4] Beyond single-photon applications, ultra-small nanodiamonds containing color centers are of great interest for bioimaging[5] and biosensing.[6,7] Among the most practical and interesting color centers in diamond are nitrogen-vacancy (NV) centers,[8] nickel-based (NE8) centers,[9] nickel-silicon complexes,[10] chromium-related color centers,[11] germanium-vacancy color centers[12] and tin-vacancy color centers,[13] as well as silicon-vacancy (SiV) centers.[14] The most developed is NV center, which has properties that include an optical readout of spin state and a long coherence time up to 1.8 ms,[15] which are the focus of interest for this center.[3] NV centers are available in single-crystal diamonds[16] and nanodiamonds[17,18] and can be produced easily during the growth process or by ion implantation[19]. Huge progress in quantum information processing has been achieved with NV centers: two-qubit quantum gates,[20] quantum register based on NV spins,[21] quantum memory at room temperature exceeding one second,[22] and much more. The NV center still has some drawbacks that limit its application, however. First, about 3% of its emission is concentrated in the zero-phonon line (ZPL)[23] while the overall spectra is 200 nm wide. In addition, the structure of the NV center is similar to the polar molecule, having high polarizability, making the NV center very sensitive to surrounding defects.[24] While this property gives rise to various sensor applications of NV center,[7,25,26] it considerably complicates any quantum information applications.

Recently, SiV centers were suggested as an alternative to NV centers[27–30] due to their strong ZPL containing about 70% of emission[31] and narrow width of around 5 nm at room temperature[14]. Another advantage is their emission at 738 nm in a spectral region where the background fluorescence of the surrounding material is weak. The first single-photon experiments on SiV centers were conducted with color centers created by ion implantation in IIa-type diamonds;[14] these, however, revealed unfavourably low, single-photon emission rates on the order of only 1000 counts/s, despite their short lifetime of 1.2 ns. In the case of nanodiamonds or shallow implanted color centers, the luminescence of SiV centers may be quenched by the presence of strain[32] or other impurities,[33] thus underscoring the need for high-purity diamonds. Recently, implanted SiV in pure diamond films has shown reasonable count rates on the order of 200 kc/s, comparable with other color centers in diamond.[34] Similar results were obtained in as-grown SiV center in bulk plates.[35] Unfortunately, even in these bright SiV centers, the radiative quantum yield of SiV color centers at room temperature was still relatively low. For example, in recent experiments for SiV centers in CVD films, the radiative quantum yield was found to be only 0.05 due to the presence of decay via optical phonons,[36] which make it difficult to use directly in quantum information application and requires strong enhancement overcoming low radiative yield.[37]

An important parameter that largely defines the application of the color center in diamond is the size of the crystal. Applications as biosensing and bioimaging require molecular-size resolution[5] that powerfully constrains nanodiamond size. Furthermore, quantum information applications can benefit a lot from ultra-small diamonds with a single SiV center inside. Here, the small size helps prevent scattering of light on the nanodiamond containing the SiV center and suppresses the phonons that cause decoherence. Ultra-small nanodiamonds containing SiV centers were successfully found in meteorite nanodiamonds of ultra-small (below 5 nm) size.[38–40] Chemical vapor deposition demonstrated possibility of creation of sub 10 nm nanodiamond, containing SiV color centers, but exhibiting some agglomeration[41]. Just recently considerable advance was reached by combining chemical vapor deposition with oxygen etching.[42] The method enabled production of sub 10-nm diamonds, contacting around 3 color centers per nanocrystal with 6.4 nm ZPL. In this paper, we demonstrate the growing of ultra-small (about 10 nm) nanodiamonds containing a single or a few bright SiV color centers with spectrally narrow ZPL line (down to 4.6 nm for single color center and on average 5.9 nm per crystal) using the high-pressure, high-temperature method (HPHT).

**Results and discussion**

The SiV center consists of a silicon atom and a lattice vacancy. The silicon atom, which substitutes a carbon atom, relaxes its lattice position towards two neighboring vacancy. This composition existed along [111][43] crystallographic axis in diamond (see Figure 1a). The defect is negatively charged and belongs to the $D_{3d}$ group of symmetry. Studied nanodiamonds with SiV color centers were synthesized by a metal-free HPHT method on the basis of a hydrocarbon-containing growth system.[44,45,46] Naphthalene $C_{10}H_8$ (Chemapol) and tetrakis (trimethylsilyl) silane $C_{12}H_{36}Si_5$ (Stream Chemicals) were used as the main hydrocarbon and silicon-doping components of the growth mixtures. Cold-pressed pellets of the initial homogeneous mixture (5 mm in diameter and 3 mm in height) were inserted into a graphite container that also served as a heater. HPHT treatment of the samples was carried out in a high-pressure, "Toroid"-type apparatus[47] (see Figure 1b). The experimental procedure consisted of loading the high-pressure apparatus to 8.0 GPa and heating the samples up to temperature of diamond synthesis (1300-1400°C) with short (about 1s) isothermal exposure at this temperature. X-ray diffraction (XRD) and Raman spectroscopy (see Supplementary Information for details), and transmission (TEM) electron microscopies were used for preliminary characterization of synthesized diamond materials. According to the obtained data, the size fraction of the diamond (ultra-small - about 10 nm, nano - up to 100 nm, submicron – 100-1000 nm and micron- more than 1000 nm ) in the conversion products is determined basically by the composition of initial growth mixture, temperature and synthesis time[45]. Adding fluorocarbon compounds (fluorinated graphite $CF_{1.1}$) to the initial growth mixtures decrease the threshold of the diamond formation temperature[44]. Therefore, synthesis at the closest point to the temperature threshold of the beginning of the formation of diamond and short (about 1 s) time of isothermal exposure led to the increase of the ultra-small fraction of diamond material in the conversion products, which can reach 50% of the total yield. In this case, taking into account the radial temperature gradient in the reaction zone of the high-pressure apparatus caused with the use of a cylindrical electrical resistivity graphite heater, the maximum content of an ultra-small sized diamond fraction is observed in the middle of the reaction zone (see Figure 2a). The maximum temperature reaches in the zone of contact of the material with the walls of a graphite heater, therefore this area is characterized by a high content of monocrystals with submicron and micron-size fractions.

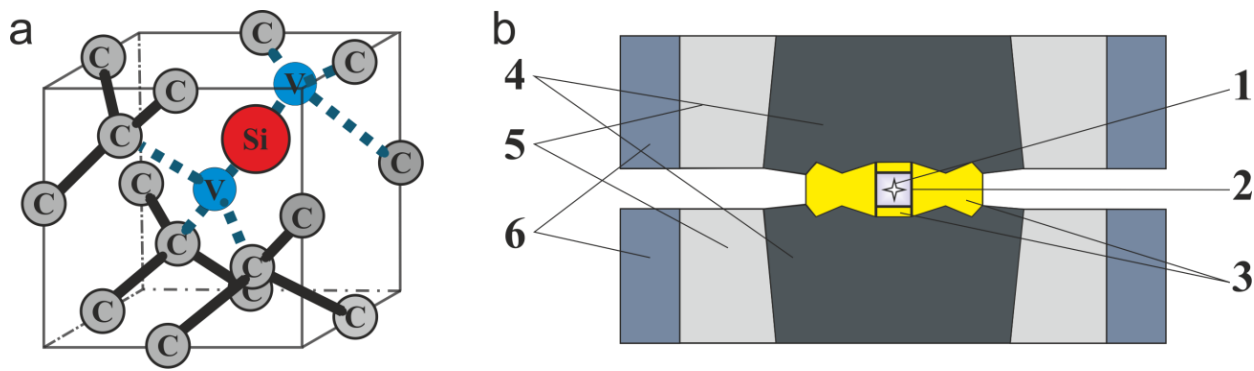

**Figure 1**. (a) The structure of the SiV center with a silicon atom in a split vacancy site. (b) Schematic depiction of an assembly of Toroid-type high pressure device used for synthesis of nano-size diamond materials on the basis of the hydrocarbon metal catalyst-free growth systems: 1 – reaction zone, 2 – graphite heater, 3 – catlinite container and catlinite thermo-insulating disks, 4 – tungsten carbide anvils, 5,6 – steel supporting rings.

The scanning electron microscope (SEM) image of the diamonds formed in the central part of the reaction zone is presented in Figure 2a. One can see a large amount of nanodiamonds (with sizes well below 100 nm) with a considerable fraction of ultra-small (<10 nm and about 10 nm) diamonds and a small quantity non-diamond, carbon materials.

Ultra-small fraction was separated using the following procedure. First, the samples were treated in a 40% solution of hydrogen peroxide for chemical purification of the diamond material. Then, separation of the ultra-small fraction of diamond was carried out in several stages that consisted of ultrasonic dispersing of diamond particles in an aqueous medium in a UP200Ht dispersant (Hielscher Ultrasonic Technology) for 1 hour at a power of 200 watts and subsequent three-stage centrifugation of aqueous dispersions of diamond materials at 1000, 3000 and 5000 g for 5 minutes each time. As a result, the diamond fraction with the dominant content of particles of about 10 nm was obtained. A Transmission Electron Microscope (TEM) image of this diamond fraction (see Figure 2c), High Resolution TEM image of the single nanodiamond (see Figure 2d) and histogram of the size distribution calculated from this are presented in Figure 2b (totally 149 nanodiamodns). The ultra-small nanodiamonds fraction was retained in the aqueous dispersion after centrifugation.

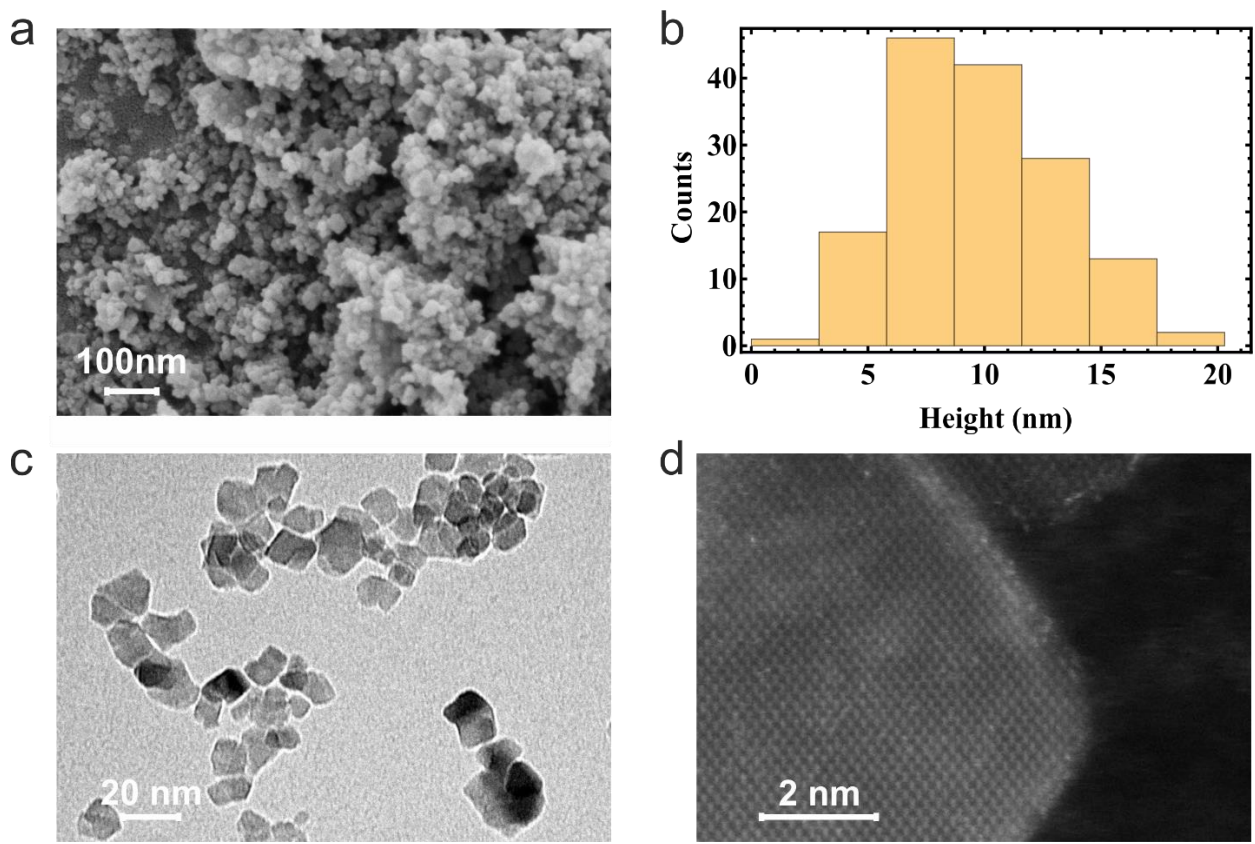

**Figure 2**. (a) SEM image of the diamond material in the middle of the reaction zone right after growing. (b) Histogram of the size distribution of synthesized nanodiamonds obtained from TEM measurements. (c)TEM image of diamond slurry. (d) HRTEM image of the single nanodiamond.

One of the interesting features of ultra-small nanodiamonds is their ability to form aggregates. The use of water as a basis for nanodiamond slurry is known to lead to aggregation during the seeding procedure[48]. To avoid this, we used a vacuum evaporation seeding method[49] (detailed in Experimental Section), which allowed us prevent aggregation during evaporation and observe SiV centers in single nanodiamonds of ultra-small sizes.

To get insight into the size of the seeded aggregates containing optically active color centers we used pre-printed gold features on the glass coverslips under investigation (see Figure 3). This allowed us to correlate images taken with a confocal microscope (see Figure 3a) and with an Atomic Force Microscope (AFM) (see Figure 3b, see[50] for more details). The AFM images were taken using an NTEGRA Spectra-M setup with silicon cantilever (NT-MDT HA_NC) operated in taping mode. In total, 3 AFM maps 10×10 μm$^2$ (256x256 points scan with lateral resolution about 40 nm and resolution in the normal direction to the scanned surface was 0.1 nm) were correlated with confocal maps. The measured sizes of the nanodiamond having color centers are presented in Figure 3c,d. From this measurement, it was verified that the distribution of nanodiamonds containing the SiV color centers (see Figure 3c) has median 9.3±2.5 nm, which is very similar to the distribution of the nanodiamond sizes themselves (see Figure 2b), whose median is 7.5±5 nm. Besides, the probability to find at least one SiV center in single nanodiamond is about 0.5%, whereas the probability to find single SiV center in single nanodiamond is approximately 0.15%. Thereby, probability to find nanodiamond with single SiV is approximately 30% among nanodiamonds with SiV color centers (see Supplementary Information for more details).Thus, color centers were observed in single nanodiamonds or just few nanodiamonds per aggregate.

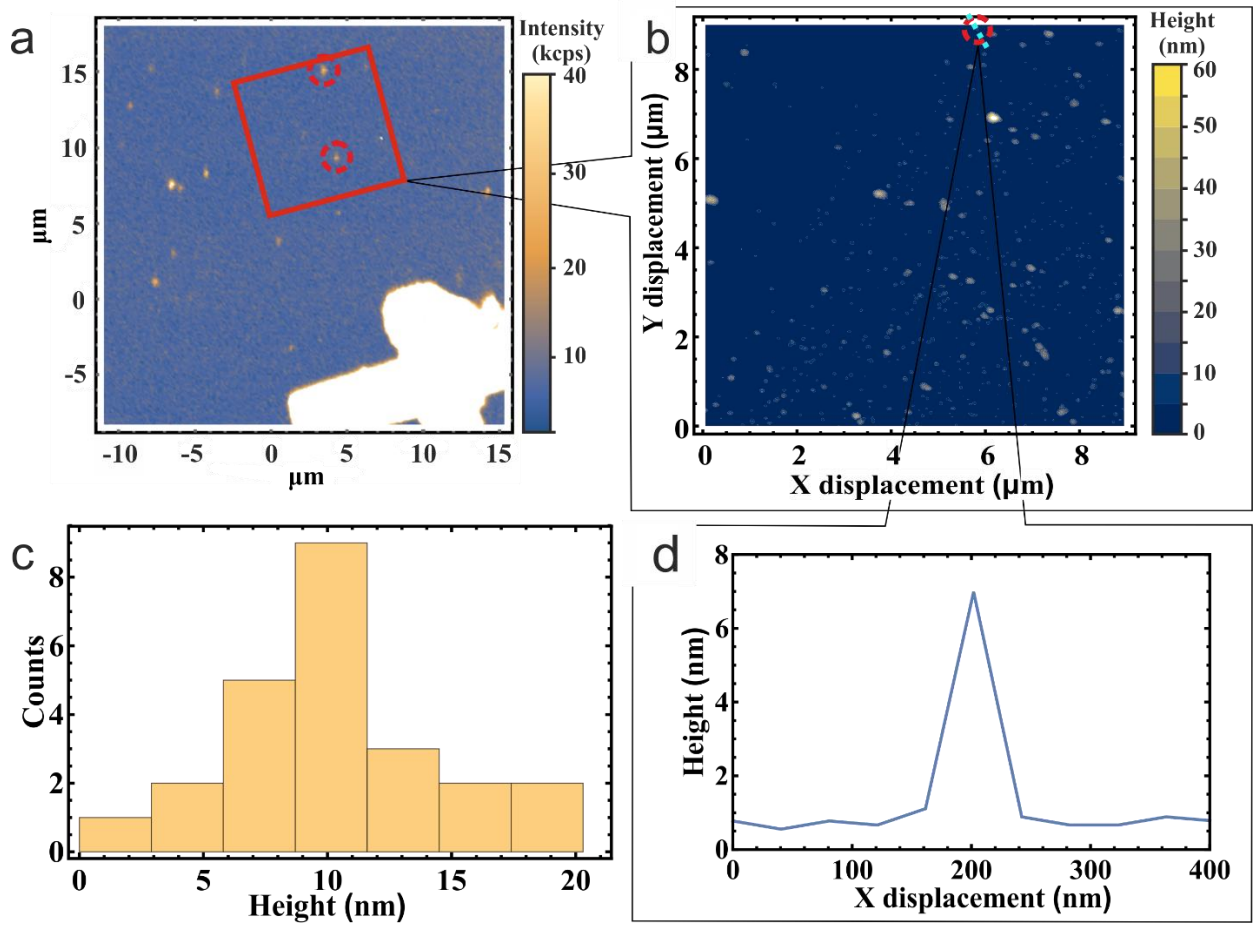

**Figure 3.** (a) Confocal map of the sample with deposited nanodiamonds. Red circles indicate nanodiamonds with optical active SiV centers. (b) AFM image of the area, marked with red square in confocal image. The blue line demonstrates direction along which 1D image was taken to determine particle size. (c) Size distribution of the nanodiamonds containing optical active SiV centers. (d) 1D profile of nanodiamond aiding in the visualization of nanodiamond size.

The optical properties of the nanodiamonds were studied using a home-built, confocal microscope with an immersion oil objective of NA 1.49. For the excitation source, we used a 532 nm diode laser (see Experimental Section for details). The relative brightness of the nanocrystals most likely is related to the number of color centers per nanocrystal. To prove that we could grow nanocrystals containing a single color center with this method, we performed second-order correlation function $g_2(\tau)$ measurements (see Figure 4a). The depth of $g_2(0)$ characterizes the number of emitters detected, while the width of $g_2(\tau)$ is limited by the color center's excited state lifetime and can only decrease as excitation power increases.[51] For nanocrystals, we measured the lifetime of 1.41 ns (see Figure 6a and 6b), while the time resolution of a pair of our detectors was measured as 672 ps (see inset in Figure 4a). Therefore, the time resolution of the detecting system considerably limits the possible depth of $g_2(\tau)$ at 0 time delay $\tau$.[51] To correct for this distortion, we performed a deconvolution procedure (see Figure 4a and Supplementary Information for more details). After deconvolution, we obtained the achieved $g_2(0)$ value, 0.41, which resulted to be below the threshold of 0.5 for two equally bright color centers.

Color center brightness depends strongly on the efficiency of excitation reaching maximum value at saturation of the center. The absorption maximum for a SiV color center is at ZPL, making it relatively hard to reach experimentally complete saturation of the center emission with green excitation.[31] Nevertheless, information on saturation counts could be extracted from analysis of the saturation curve (Figure 4b). To analyze the saturation curve, we used the standard model for fitting:

$$I = \frac{I_0 \times p}{p_0 + p} + \alpha p, \quad (1)$$

where $I_0$ – counts in saturation, $p$ – laser power, $p_0$ – laser power in saturation, $\alpha$ – coefficient for a linear substrate noise. Here, we analyzed the emission of 7 single color centers in our diamonds. By fitting these experimental data, we found that single SiV center in nanocrystal has average saturation count rate of 75±47 kcounts/s at the saturation power 3.2±2.4 mW; this values were gained from fit's extrapolation because reaching the saturation with green laser can be relatively hard. The whole error is square root from statistical measurements and fit error. The observed saturation behavior greatly varied from center to center (Figure 4b), with the fitted $I_0$ values ranging from 19±6 kcounts/s to 143±30 kcounts/s, and the $p_0$ ranging from 0.6±0.3 mW to 6.9±2.3 mW. This intensity level is comparable with SiV centers in CVD nanodiamonds,[52] HPHT nanodiamonds,[53] as grown in bulk diamond plate[35] and ion implantation followed by annealing.[34] Remarkably, this high count rate is accompanied by rather stable emission, depicted in Figure 4c.

In addition, we performed detailed spectral studies for various nanodiamonds ranging from nanocrystals containing a single SiV to nanocrystals or aggregates containing ensembles with approximately 10 SiV centers. The approximate number of color centers was estimated as ration of saturation count to the single color center average number of counts in saturation.

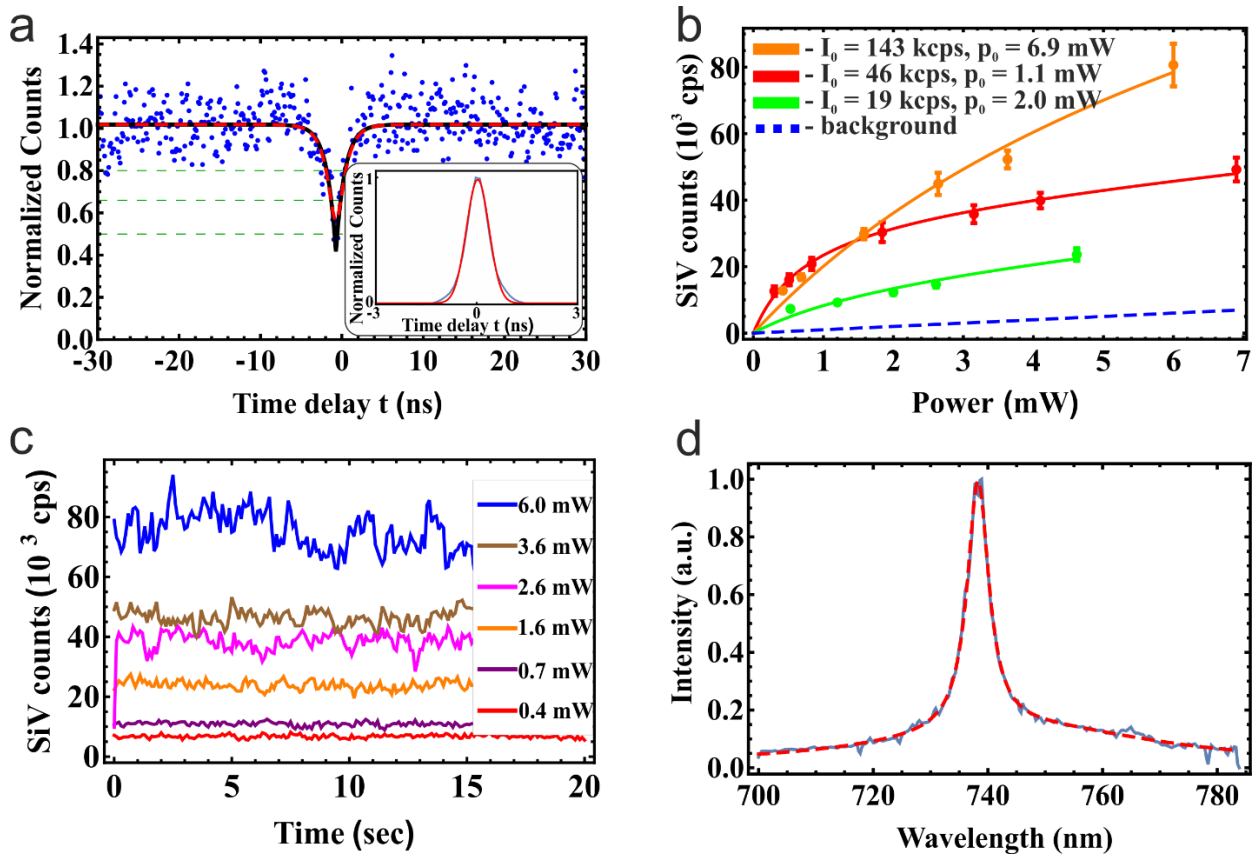

**Figure 4.** (a) Second-order autocorrelation function obtained from a single SiV center. The inset illustrates measurement of detectors' jitter and its approximation by Gaussian distribution. The green horizontal lines show $g_2(0)$ for 2, 3 and 4 equally bright emitters, assuming zero electronic noise level. (b) Saturation curves for a single SiV centers. The red, orange and green solid lines represent the fit to the saturation; the dashed blue line is the background contribution to the counts detected. (c) Time trace of SiV center under various excitation powers. (d) Spectra obtained for a single SiV center, FWHM = 4.6 nm.

Figure 4d shows the typical spectrum of SiV centers in synthesized nanodiamonds. We measured a total of 24 spectrums for SiV centers, contained in nanocrystal found from correlated confocal and AFM maps. Besides spectrum of extra 11 nanocrystals was recorded, for which correlation measurement were not performed. Each spectrum was fitted with the sum of 3 Lorentzian functions. One of the Lorenzians approximated ZPL and the rest, phonon sideband. We found that the ZPL position has a narrow distribution around 738.06 nm, with a standard deviation of 0.27 nm (Figure 5a). This narrow distribution indicates the low level of strain in our nanodiamond, because ZPL line position in the general depends on the strain.[54] We note that the spread of the ZPL line position for SiV centers grown by other methods is considerably higher, as previously indicated: in the case of CVD nanodiamond synthesis[55] it is about 1.5 nm. For HPHT nanodiamonds grown with other conditions ZPL line position is 737 ± 10 nm.[53] For ultra-small nanodiamonds found in natural meteorite, ZPL position was found to be 735.7 with variations of ±5 nm.[38] At the same time, the width of ZPL line at room temperature has a mean of 5.9 nm with a standard deviation of 0.8 nm (Figure 5b). This is a rather standard value determined mostly by the strong coupling of excited levels of SiV color centers via the phonons in diamond.[56] This mixing is obviously temperature depended[57] and thus should only be compared at known temperature. Raman spectroscopy for diamond material before and after cleaning procedure allow us made assessments of the strain in synthesized nanodiamonds using changes in diamond Raman line position[58,59] as well as bonds hybridization[60], which is mainly sp3 for our diamonds (see Supplementary information for details). On the other hand, analysis the ZPL line position statistics, which depends on strain, allow us independently estimate the strain. Both methods are in a good agreement with each other and allow to conclude, that the strain level in produced nanodiamonds less than 0.7 GPa (see Supplementary Information for details).

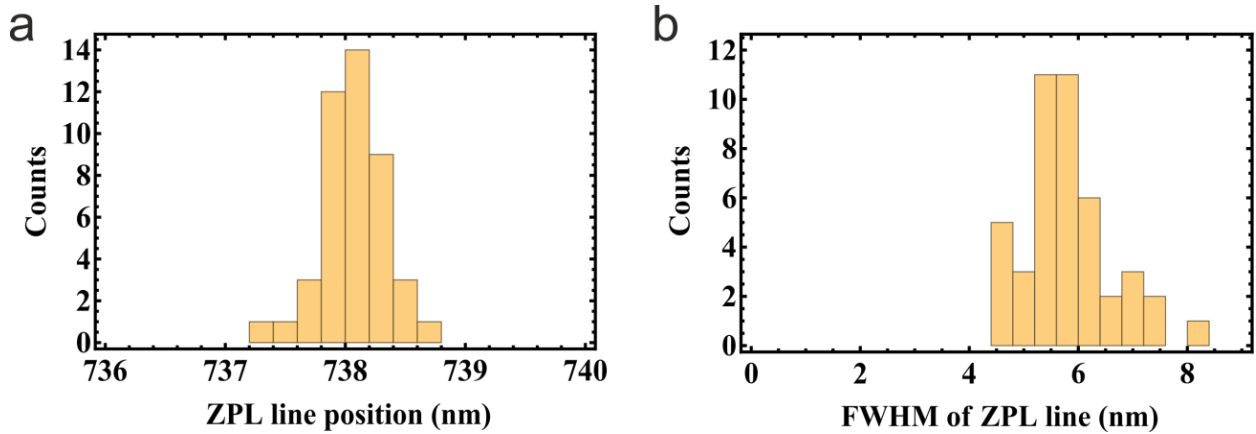

**Figure 5.** (a) Histogram of ZPL line position. (b) Histogram of FWHM of ZPL lines of SiV centers observed from 1–10SiV centers per crystal.

**Conclusions**

We synthesized ultra-small nanodiamonds (smaller than 10 nm) containing single or few SiV color centers. The nanodiamonds demonstrate high single-photon brightness (>100 000 cps) at room temperature. Both the nanocrystals containing a single color center and those with relatively large concentrations of SiV centers (up to 10 SiV per 10 nm crystal) have narrowband luminescence and extremely stable ZPL position, which evidence strain level less than 0.7 GPa and correspondingly low number of defects, different from SiV centers s in the nanodiamonds grown. This high quality, ultra-small nanodiamond may be of great interest for quantum information and bioimaging applications.

**Experimental section**

**Sample preparation**

Ultra-small nanodiamonds have a strong tendency toward aggregation, and in order to prevent clustering, we performed the following procedure. First, the glass coverslip (Menzel Gläser) with a preexisting gold mask was cleaned in a piranha solution ($H_2SO_4$ and $H_2O_2$ in a ratio of 3 to 1 respectively), and kept at $120°C$ for 1 hour. Then, the slurry with nanodiamonds was sonicated for 1 hour at a power of 170 watts. For sonication the Ultrasonic Cleaner CD-4820 was used. Finally, the 10 μl of slurry was dropped onto the coverslip and evaporated in the vacuum chamber using a Harrick Plasma PDC-002 plasma cleaner.[49]

**Confocal setup**

To analyze the optical properties of SiV centers, we used a home-built confocal setup (Figure 6a) with a pumping diode laser at 532 nm (Coherent Compass 300) for exciting the SiV center in nanodiamonds, and galvo mirrors (Cambridge Technologies) for scanning the specimen. The emission was collected with a high numerical aperture oil objective (NIKON Apo TIRF 100X NA 1.49) with a working distance of 120 micrometers. The assembled fluorescence passes through the combination of the longpass optical filter with cut-off at 600 nm and notch filter with a stop-band centered at 532 nm to reject the excitation light and Raman signal from the collected emission. The luminescence that passed through the filters was coupled into a single-mode fiber (Thorlabs SM600) and guided to a Hunbury-Brown-Twiss (HBT) interferometer that consisted of two avalanche photodiodes (PerkinElmer SPCM-AQRH-14-FC) and a 45:55 beam splitter, or directed to the home-built spectrometer with monochromator (Jarrell Ash model 82-422) and CCD camera (Starlight Xpress SXV-H9C). A time-correlated, single-photon counting module (Picoquant Picoharp 300) was used to obtain second-order photon correlation histograms from SiV centers. For the lifetime measurements, a picosecond diode laser (Picoquant LDH-P-FA-530 XL) was used with a same time-correlation, single-photon counting module. Due to the fact that the lifetime of the SiV center excited state is longer than the length of the laser pulse, the fluorescence signal recorded after the applied laser pulse should exhibit exponential decay. In fact, the actual signal observed consisted of two exponents: fast and slow (Figure 6b and Figure 6c). If the fast exponential component corresponds to background fluorescence signatures that do not have a slowly decaying exponential component, then the slow decay "tail" is associated with the lifetime of SiV center in an excited state.

With the exception of the spectrometer measurements, the emission was additionally filtered by a 737±10 nm narrowband filter.

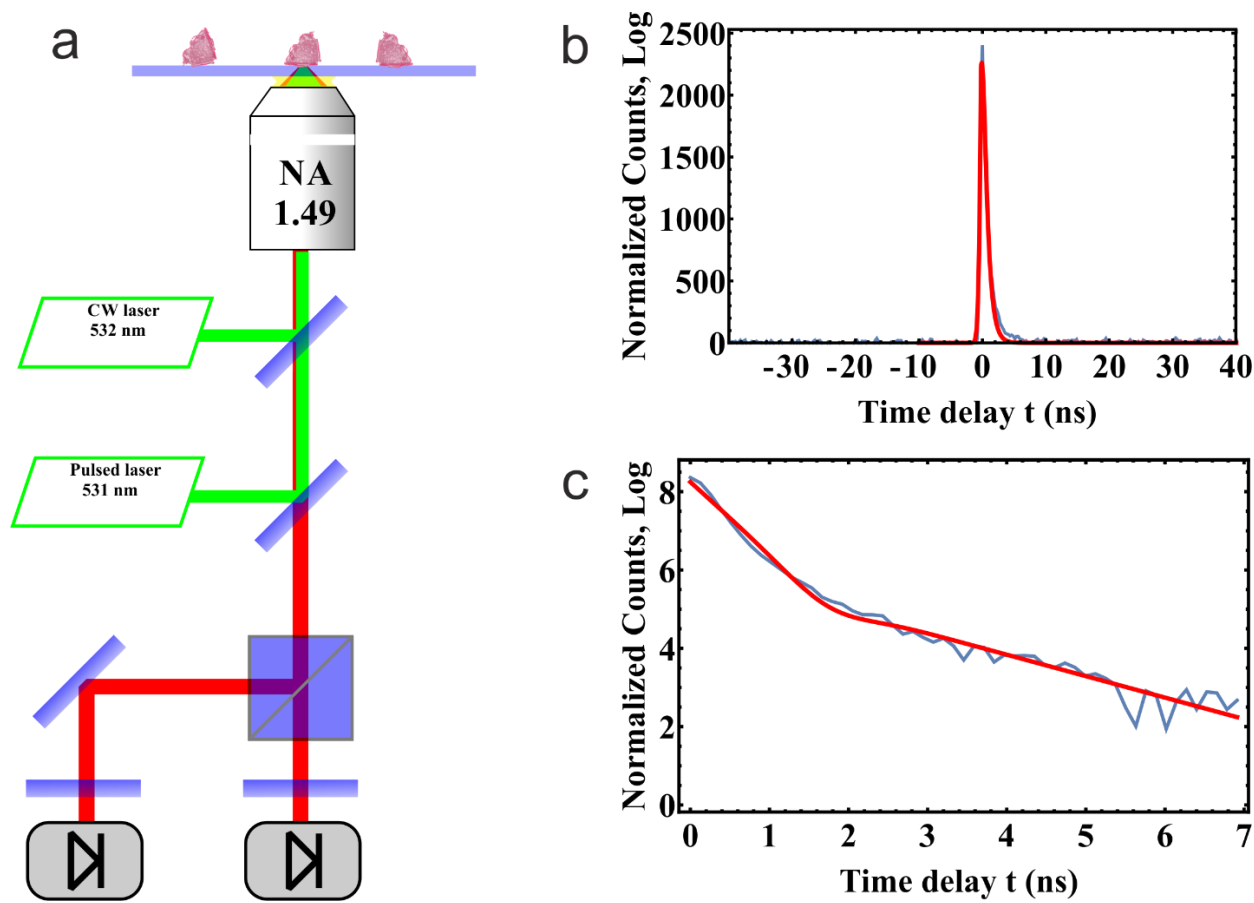

**Figure 6.** (a) Schematic of the confocal microscope. (b-c) Lifetime measurement for SiV center. Fast decay during the first nanoseconds is due to background fluorescence.

ASSOCIATED CONTENT

Calculation of probability to finding SiV in single nanodiamonds, size measurements from AFM maps, Raman measurements and preliminary characterization of grown diamond material, strain assessment by the ZPL position analysis, measurement of the setup jitter function and deconvolution procedure described in the supporting information (PDF).

AUTHOR INFORMATION

**Corresponding Author**

Alexey Akimov

*E-mail: aa@rqc.ru

**Author Contributions**

The main contributors of this paper are Stepan Bolshedvorskii and Anton Zeleneev, who performed majority of the optical measurements, developed and realize correlation of confocal and AFM maps, and Valery Davydov who was responsible for diamond synthesis. Alexey Akimov developed the initial idea of this work and supervised all parts of the study. Vadim Vorobyov, Vladimir Soshenko and Olga Rubinas helped with acquiring, analysis of optical data, Leonid Zhulikov prepared the samples, and analyzed nanodiamonds positioning on the sample, Pavel Pivovarov performed AFM measurements. Vadim Sorokin and Andrey Smolyaninov supervised optical and AFM investigation process. Valery Davydov and Liudmila Kulikova performed full sample diamond material synthesis and purification procedure. Sergey Lyapin performed Raman measurements of full sample diamond material. Anastasia Granina, Viatcheslav Agafonov and Rustem Uzbekov performed separation of the ultra-small-sized fraction of diamonds and Transmission Electron Microscope investigation of the diamond materials using for optical experiments.

**Funding Sources**

This work was supported by the Russian Science Foundation (Grant 19-19-00693) in part of the sample analysis. Sample fabrication was supported by the Russian Foundation for Basic Research (Grant No. 18-03-00936).

**Notes**

The authors declare no competing financial interest.

ACKNOWLEDGMENT

This work was supported by the Russian Science Foundation (Grant 19-19-00693) in part of the sample analysis. Sample fabrication was supported by the Russian Foundation for Basic Research (Grant No. 18-03-00936).

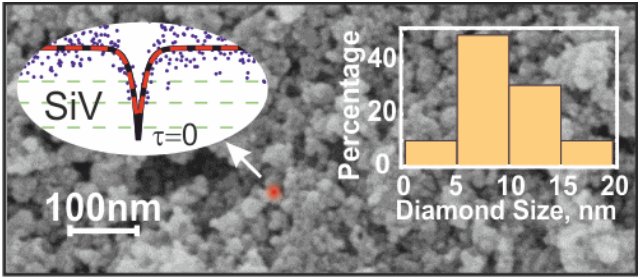